\documentclass[10pt,aps,pre,twocolumn,footinbib,tightenlines,superscriptaddress,final,letterpaper,longbibliography,floatfix]{revtex4-2}

\usepackage{silence}
\usepackage[T1]{fontenc}
\usepackage[utf8]{inputenc}

\usepackage{amsmath, dsfont, amsfonts}
\usepackage{physics}
\usepackage{hyperref}
\usepackage{orcidlink}
\usepackage{graphicx, float}
\usepackage[caption=false]{subfig}
\usepackage{xcolor}

\definecolor{monbbleu}{RGB}{76, 114, 176}
\hypersetup{unicode=false, pdftoolbar=true, pdfmenubar=true, pdffitwindow=false, pdfstartview={FitH}, pdfnewwindow=true, pdftitle={}, pdfauthor={}, pdfsubject={}, pdfkeywords={}, colorlinks=true, allcolors=monbbleu}

\makeatletter
\newcommand\thefontsize{The current font size is: \f@size pt}
\makeatother

\newcommand{\prob}{\mathbb{P}}

\begin{document}

\title{Symmetry-driven embedding of networks in hyperbolic space}

\author{Simon Lizotte\,\orcidlink{0000-0002-5446-675X}\,}
\affiliation{D\'epartement de physique, de g\'enie physique et d'optique, Universit\'e Laval, Qu\'ebec (Qu\'ebec), Canada G1V 0A6}%
\affiliation{Centre interdisciplinaire en mod\'elisation math\'ematique, Universit\'e Laval, Qu\'ebec (Qu\'ebec), Canada G1V 0A6}%

\author{Jean-Gabriel Young\,\orcidlink{0000-0002-4464-2692}\,}
\affiliation{D\'epartement de physique, de g\'enie physique et d'optique, Universit\'e Laval, Qu\'ebec (Qu\'ebec), Canada G1V 0A6}%
\affiliation{Department of Mathematics and Statistics, University of Vermont, Burlington, VT 05405, USA}
\affiliation{Vermont Complex Systems Center, University of Vermont, Burlington, VT 05405, USA}

\author{Antoine Allard\,\orcidlink{0000-0002-8208-9920}\,}
\affiliation{D\'epartement de physique, de g\'enie physique et d'optique, Universit\'e Laval, Qu\'ebec (Qu\'ebec), Canada G1V 0A6}%
\affiliation{Centre interdisciplinaire en mod\'elisation math\'ematique, Universit\'e Laval, Qu\'ebec (Qu\'ebec), Canada G1V 0A6}%
\affiliation{Vermont Complex Systems Center, University of Vermont, Burlington, VT 05405, USA}

\begin{abstract}
    Hyperbolic models are known to produce networks with properties observed empirically in most network datasets, including heavy-tailed degree distribution, high clustering, and hierarchical structures.
    As a result, several embeddings algorithms have been proposed to invert these models and assign hyperbolic coordinates to network data.
    Current algorithms for finding these coordinates, however, do not quantify uncertainty in the inferred coordinates.
    We present BIGUE, a Markov chain Monte Carlo (MCMC) algorithm that samples the posterior distribution of a Bayesian hyperbolic random graph model.
    We show that the samples are consistent with current algorithms while providing added credible intervals for the coordinates and all network properties.
    We also show that some networks admit two or more plausible embeddings, a feature that an optimization algorithm can easily overlook.
\end{abstract}

\maketitle

\section{Introduction}

Embedding enables us to comprehend abstract objects and conduct rigorous quantitative analyses of their similarities, differences, and structural characteristics.
In the complex network context, embeddings usually consist of vertex coordinates in a latent space.
A rich literature has shown that embedding in hyperbolic spaces~\cite{boguna_2021_NetworkGeometry}, in particular, can visually emphasize the important parts of large networks~\cite{munzner_1998_ExploringLargeGraphs}, but also, when used as the basis for a random graph model, naturally reproduce many common network properties such as the community structure~\cite{kovacs_2021_InherentCommunityStructure, foster_2011_ClusteringDrivesAssortativity, balogh_2023_MaximallyModularStructure, balogh_2024_IntracommunityLinkFormation, zuev_2015_EmergenceSoftCommunities, garcia-perez2018SoftCommunitiesSimilarity, faqeeh_2018_CharacterizingAnalogyHyperbolic}, a power-law degree distribution~\cite{krioukov_2010_HyperbolicGeometryComplex, gugelmann_2012_RandomHyperbolicGraphs}, a non-vanishing clustering coefficient~\cite{krioukov_2010_HyperbolicGeometryComplex, krioukov_2016_ClusteringImpliesGeometry, gugelmann_2012_RandomHyperbolicGraphs, candellero_2014_ClusteringHyperbolicGeometry, boguna_2020_SmallWorldsClustering, serrano_2008_SelfSimilarityComplexNetworks}, a fractal structure~\cite{zheng_2020_GeometricRenormalizationUnravels, garcia-perez_2018_MultiscaleUnfoldingReal, song_2005_SelfsimilarityComplexNetworks} and the sparsity of connections~\cite{boguna_2020_SmallWorldsClustering}.
Furthermore, since hyperbolic geometry closely approximate the shortest paths of graphs via geodesics~\cite{krioukov_2010_HyperbolicGeometryComplex, boguna_2009_NavigabilityComplexNetworks}, hyperbolic embeddings have been used to design efficient routing protocols~\cite{boguna_2010_SustainingInternetHyperbolic, papadopoulos_2010_GreedyForwardingDynamic}.

Considerable effort has gone into representing a given graph using latent spaces~\cite{baptista_2023_ZooGuideNetwork, cai_2018_ComprehensiveSurveyGraph, zhang_2021_SystematicComparisonGraph, mcdonald_2021_HierarchicalOrganisationDynamics, goyal_2018_GraphEmbeddingTechniques}.
Many studies focus on placing the vertices such that the graph distance between the vertices is closely approximated by the hyperbolic geodesics~\cite{chowdhary_2018_ImprovedHyperbolicEmbedding, keller-ressel_2020_HydraMethodStrainminimizing, sala_2018_RepresentationTradeoffsHyperbolic, verbeek_2016_MetricEmbeddingHyperbolic, clough_2017_EmbeddingGraphsLorentzian}.
From this perspective, trees~\cite{sarkar_2012_LowDistortionDelaunay} can be embedded in hyperbolic space up to an arbitrarily small error, enabling perfect greedy routing.
This is due to the exponentially growing volume in the hyperbolic space, which can be related to the branching factor of trees~\cite{krioukov_2010_HyperbolicGeometryComplex}.
The embedding task was also tackled with dimension reduction techniques from machine learning~\cite{goyal_2018_GraphEmbeddingTechniques, chamberlain_2017_NeuralEmbeddingsGraphs, cannistraci_2013_MinimumCurvilinearityEnhance, muscoloni_2017_MachineLearningMeets} and with the maximum likelihood estimation of various hyperbolic random graph models~\cite{garcia-perez_2019_MercatorUncoveringFaithful, boguna_2010_SustainingInternetHyperbolic, papadopoulos_2015_NetworkMappingReplaying, papadopoulos_2015_NetworkGeometryInference, alanis-lobato_2016_ManifoldLearningMaximum, alanis-lobato_2016_EfficientEmbeddingComplex, wang_2016_FastCommunityDetection, wang_2016_LinkPredictionBased, blasius_2018_EfficientEmbeddingScaleFree, wang_2019_FastHyperbolicMapping, ye_2022_CommunityPreservingMapping}.
Other methods allow the embedding of directed graphs~\cite{wu_2020_AsymmetricPopularitySimilarityOptimization, kovacs_2023_ModelindependentEmbeddingDirected,allard_2024_GeometricDescriptionClustering}, weighted graphs~\cite{yi_2021_HyperbolicEmbeddingMethod,allard_2017_GeometricNatureWeights} and graphs to the $D$-dimensional hyperbolic space with~\cite{jankowski_2024_FeatureawareUltralowDimensional} or without vertex features~\cite{jankowski_2023_DMercatorMethodMultidimensional}.

\begin{figure}[b]
    \centering
    \includegraphics[width=.6\columnwidth]{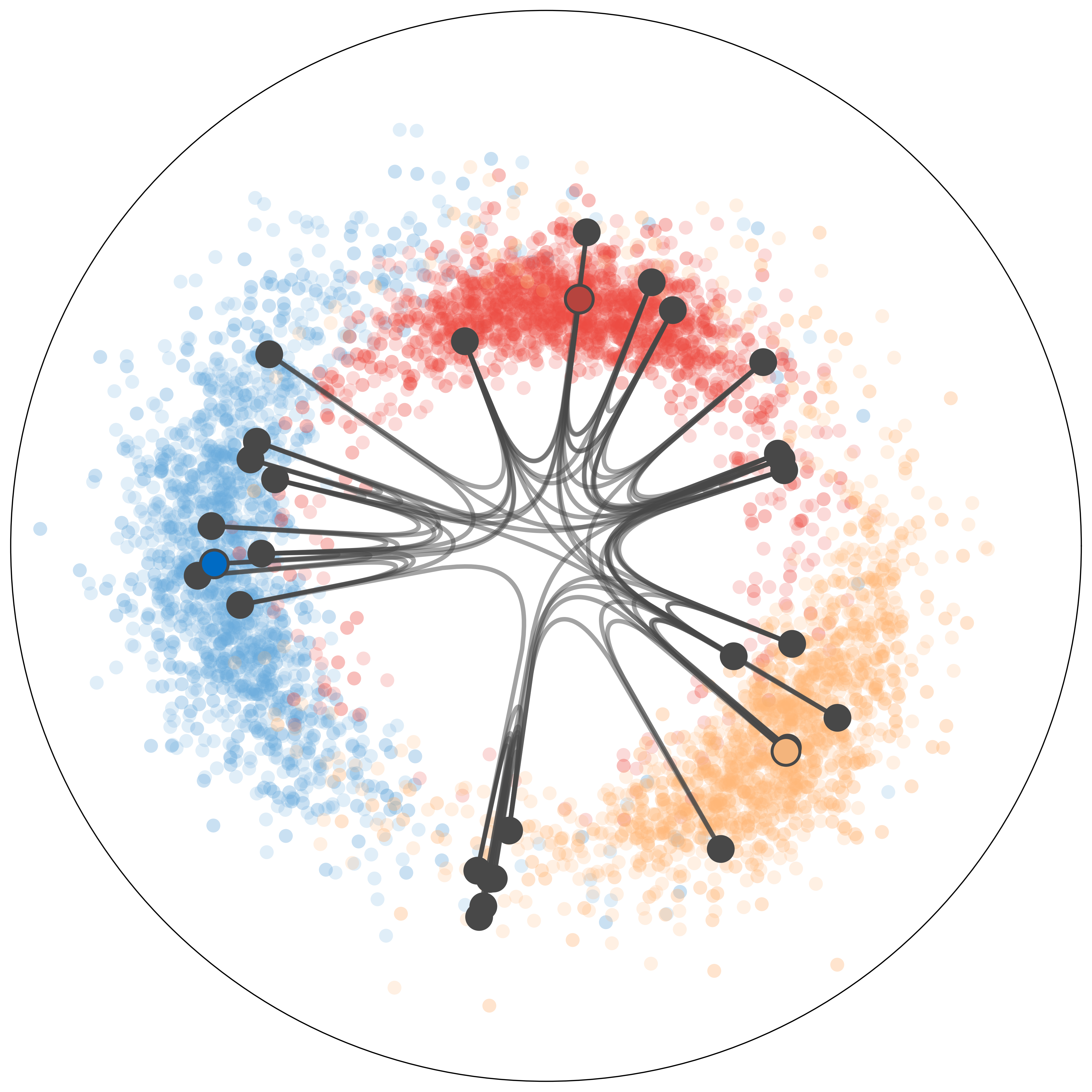}
    \caption{Probabilistic hyperbolic embedding of a synthetic graph with BIGUE.
    Hyperbolic coordinates are obtained using a transformation between the $\mathbb{H}^2$ and the $\mathbb{S}^1$ models described in Supplementary Note~2.
    Black points and dark-colored points are the median coordinates of each vertex.
    Light-colored points are the $2000$ sampled positions for the three highlighted vertices.
    Lines are edges drawn using hyperbolic geodesics.
    The synthetic graph of 30 vertices is generated with the $\mathbb{S}^1$ likelihood, with angular coordinates drawn from their prior, $\beta=2.5$ and the $\kappa$ parameters drawn from a Pareto distribution of exponent $-2.5$ over the truncated interval $(4, 10)$.
    }
    \label{fig:probabilistic_embedding}
\end{figure}

Existing methods have, however, largely overlooked uncertainty: They return a single embedding without allowance for error or acknowledging the existence of other solutions.
Yet, this information is crucial as perturbations to the embedding affect edge prediction~\cite{kitsak_2020_LinkPredictionHyperbolic} and can lead to different graph representations.

As a first step towards addressing this problem, we propose a Bayesian embedding algorithm that quantifies uncertainty rigorously.
Related methods have proven useful in network science, where they have been used to reconstruct networks from noisy measurements ~\cite{newman_2018_NetworkStructureRich, young_2021_BayesianInferenceNetwork}, predict the existence of edges~\cite{peixoto_2018_ReconstructingNetworksUnknown}, infer higher-order interactions from network data~\cite{young_2021_HypergraphReconstructionNetwork} and pairwise observation~\cite{lizotte_2023_HypergraphReconstructionUncertain}, and perform community detection~\cite{peixoto_2019_BayesianStochasticBlockmodeling, pas_2018_BayesianCommunityDetection}.

Our proposed algorithm, BIGUE (Bayesian Inference of a Graph's Unknown Embedding), uses Markov chain Monte Carlo (MCMC) sampling to generate embeddings compatible with a given graph; see Fig.~\ref{fig:probabilistic_embedding}.
We show how these samples can be used to calculate credible intervals, interpreted as error bars, for vertex positions, properties of the embedding, and model parameters.
We also show that many qualitatively different embeddings are often compatible with an observed graph---a phenomenon known as multimodality.

Bayesian approaches hyperbolic embeddings of networks have already been studied~\cite{papamichalis_2022_LatentSpaceNetwork}, but the results therein suggest that the proposed MCMC algorithm has poor mixing---a common problem many MCMC methods face when the posterior distribution is complex and multimodal~\cite{yao_2022_StackingNonmixingBayesian, yao_2020_AdaptivePathSampling, west_2022_KernelMixingStrategy, tjelmeland_2001_ModeJumpingProposals, pompe_2020_FrameworkAdaptiveMCMC, park_2021_SamplingMultimodalDistributions, neal_1996_SamplingMultimodalDistributions, lan_2014_WormholeHamiltonianMonte, graham_2017_ContinuouslyTemperedHamiltonian}.
We identify community structure~\cite{faqeeh_2018_CharacterizingAnalogyHyperbolic} as a source of multimodality and introduce a set of cluster-based transformations to improve the exploration of the posterior distribution.
Our approach draws on previous literature in which community structure was used to guide greedy embedding algorithms~\cite{wang_2019_FastHyperbolicMapping, wang_2016_HyperbolicMappingComplex}.
We show that these transformations drastically improve the mixing of the MCMC algorithm, even if the community structure is weak.
In doing so, we make a tradeoff between speed and accuracy---greedy methods find a single solution rapidly, while our MCMC methods explore a large space more slowly.
In fact, our implementation of BIGUE does not scale well beyond a few hundred vertices.
Thad said we show that the cluster transformations powering BIGUE improve mixing and thus make it more scalable than alternative MCMC algorithms.
Furthermore, we provide evidence that the posterior distributions of hyperbolic graph models are not Gaussian in practice,  thereby motivating the need for MCMC as opposed to faster but less accurate methods like variational inference.

\section{Methods}
\subsection{Bayesian hyperbolic embedding}

The Bayesian framework aims to infer parameters---here the embedding---from some data---the graph.
In this section, we define the likelihood and prior distributions that lead to the posterior of the model, and highlight unique challenges that do not arise when searching for a single embedding.

\label{ssec:bayesian_model}
\subsubsection{Model definition}
\label{ssec:model_def}

Let $G=(V, E)$ be the graph we aim to embed, with $V$ and $E$ being the sets of its vertices and of its edges, respectively.
Let $\mathcal{G}$ be the random graph used to perform the embedding.
We use the circular probabilistic model $\mathbb{S}^1$ to describe the graph $G$~\cite{serrano_2008_SelfSimilarityComplexNetworks}; it is nearly equivalent to the hyperbolic plane model $\mathbb{H}^2$~\cite{krioukov_2010_HyperbolicGeometryComplex} (see Supplementary Note~2), but it will turn out to facilitate inference.
In the $\mathbb{S}^1$ model, the probability that an edge $(u,v)$ exists between vertices $u$ and $v$ is a function of the angular coordinates $\theta = (\theta_w)_{w\in V}$ of the vertices where each $\theta_w \in [-\pi, \pi)$, of the parameters $\kappa = (\kappa_w)_{w\in V}$ where $\kappa_w>0$ controls the degree of vertex $w$, and of an inverse temperature $\beta>1$ which controls the amount of clustering.
This probability is defined as
\begin{align}
    \prob[a_{uv} = 1 | \theta, \kappa, \beta] = \qty(1+\qty(\frac{d(\theta_u, \theta_v)}{\mu \kappa_u \kappa_v})^\beta)^{-1},
    \label{eq:edgeprob}
\end{align}
where $a_{uv}=a_{vu}$, $u\neq v$, is any element of the adjacency matrix of a graph realized by $\mathcal{G}$, $d(\theta_u, \theta_v)=R\Delta(\theta_u, \theta_v)$ is the arc length between vertices $u$ and $v$, $\Delta(\theta_u,\theta_v) = \pi-|\pi-|\theta_u - \theta_v||$ is the angular separation, and $R=|V|/2\pi$ is the radius of the circle on which vertices are embedded.
We fix the constant $\mu = \beta \sin(\pi/\beta)/ (2\pi \mathbb{E}[\tilde{\kappa}])$ so that the parameters $\kappa$ match the degrees $\mathbb{E}[K | \kappa]=\kappa$ in expectation in the limit of large graphs, where  $\tilde{\kappa}$ represents the random variable for the parameter $\kappa_w$ of any vertex $w$ (they are i.i.d.) and $K|\kappa$ is the random degree of a vertex of parameter $\kappa$ when all vertices are independently and uniformly distributed on the circle~\cite{krioukov_2010_HyperbolicGeometryComplex}.
These choices for $R$ and $\mu$ are without lost of generality.

The complete graph is modeled using conditional independence for the edges, leading to the likelihood
\begin{align}
    \label{eq:likelihood}
    \prob[\mathcal{G}=G | \theta, \kappa, \beta] &=
    \prod_{(u,v) \in E} \prob[a_{uv} = 1 | \theta, \kappa, \beta] \nonumber \\
    &\hspace{10pt}\times \prod_{(u,v) \not \in E} (1-\prob[a_{uv} = 1 | \theta, \kappa, \beta]) \nonumber \\
    &= \prod_{(u,v) \in \binom{V}{2}} \qty(1+\qty(\frac{d(\theta_u, \theta_v)}{\mu \kappa_u \kappa_v})^{\beta_{uv}^\pm })^{-1},
\end{align}
where $\binom{V}{2}$ is the set of all combinations of two vertices, and $\beta_{uv}^\pm$ equals $\beta$ if $a_{uv}=1$ and equals $-\beta$ otherwise.

Applying Bayes' rule yields the following posterior density  for the embedding (i.e., angular coordinates and parameters)
\begin{align}
    p(\theta, \kappa, \beta | G) = \frac{\prob[\mathcal{G}=G | \theta, \kappa, \beta] p(\theta, \kappa, \beta)}{\prob[\mathcal{G}=G]}.
    \label{eq:bayesian_model}
\end{align}
The likelihood $\prob[\mathcal{G}=G | \theta, \kappa, \beta]$ is given by  Eq.~\eqref{eq:likelihood}.
To complete the specification of Eq.~\eqref{eq:bayesian_model}, we use independent prior densities
\begin{align}
    p(\theta, \kappa, \beta) = p(\beta) \prod_{w\in V} p(k_w) p(\theta_w),
\end{align}
for ease of modeling, though our sampling algorithm does not depend on this simplifying assumption.
Previous literature has shown that inverse temperatures are typically small; we thus choose a truncated half-normal distribution for $\beta$,
\begin{align}
    p(\beta) \propto \mathds{1}[\beta>1] \exp{-\frac{(\beta-\beta_0)^2}{2\sigma^2}}
\end{align}
with $\beta_0=3$ and $\sigma=2$, and where $\mathds{1}[\cdot]$ is the indicator function (equals $1$ is its argument is true, and $0$ otherwise).
Since known networks tend to have a heavy-tailed degree distribution, we use a half-Cauchy prior
\begin{align}
    p(\kappa_w) \propto \mathds{1}[\kappa_w>\varepsilon] \frac{2}{\pi\gamma\qty(1+(\kappa_w/\gamma)^2)}, \quad \forall w\in V
    \label{eq:kappa_prior}
\end{align}
with $\gamma=4$ and $\varepsilon = 10^{-10}$; this allows for large variations in the estimated values of $\kappa$.
Finally, because the angular coordinates are defined on a bounded space and we do not want to favor an embedding direction a priori, we use a uniform prior
\begin{align}
    \label{eq:theta_prior}
    p(\theta_w) = \frac{1}{2\pi}
\end{align}
except for the two vertices with the highest degree, whose position is restricted (the reason is technical and has to do with avoiding degeneracy in the posterior; see below).
With these choices, the posterior distribution is proper and can express correlations between the parameters even if the priors are independent.
The normalization constant $\prob[\mathcal{G}=G]$ can be obtained by integration in principle, though it is not needed here; see the Sampling section below.

\begin{figure}
    \centering
    \includegraphics{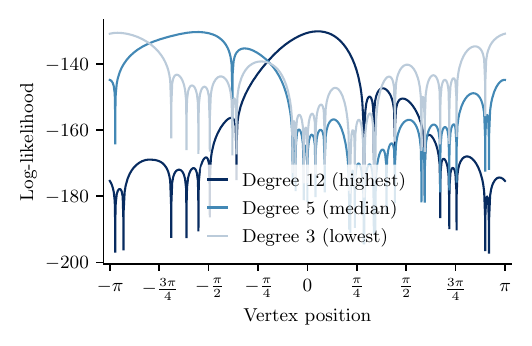}
    \caption{
        Log-likelihood of the model when a single vertex is moved along the circle.
        Each dip is a divergence that occurs when a vertex is positioned at the same position as a disconnected vertex.
        These divergences form barriers in the objective function landscape (log-likelihood, posterior distribution), which is one of the reasons why hyperbolic embedding is a difficult problem.
        The likelihood is computed on the ground truth embedding of the synthetic graph of Fig.~\ref{fig:probabilistic_embedding}.
        Colors indicate the degree of the moved vertex (lowest degree in light blue, median degree in blue and highest degree in dark blue).
    }
    \label{fig:vertex_loglikelihood}
\end{figure}

\subsubsection{Symmetries and automorphisms}
\label{subsec:sym_auth}
The $\mathbb{S}^1$ model exhibits fundamental symmetries that give rise to identifiability problems.
While these symmetries are often overlooked when using maximization algorithms (which produce single solutions), they must be addressed in the Bayesian context.
Indeed, our goal is to compute a distribution over embeddings, requiring us to distinguish between embeddings that differ merely by trivial reparametrizations (such as coordinate rotations) and those that represent genuinely distinct configurations.
In this section, we examine the two classes of symmetries inherent in the model and present our approach for handling them.

First, the model inherits the symmetries of the circle $\mathbb{S}^1$.
Since the distance $d$ is invariant to rotations and reflections $\theta_w \mapsto -\theta_w \ \forall w$, the edge probabilities are preserved in the likelihood, and an infinite number of equivalent embeddings exist.
Maximization algorithms break this symmetry at random, but we need to proceed with more caution when we move to a distribution over embedding.
Without loss of generality, we fix the angular coordinate of the highest-degree vertex $u$ to $\theta_u=0$ to handle rotational symmetries.
We remove the reflection symmetry by restricting the angular coordinate of the second highest-degree vertex $v$ to $\theta_v \in [0, \pi]$.

Second, a graph is inherently labeled by its vertices, but any vertex relabelling that preserves the edges---an automorphism---leaves the likelihood invariant since the distances between connected and unconnected vertices are unchanged.
For example, if a transformation $u\leftrightarrow v$ exists such that each $(u,w)\in E \Longleftrightarrow (v, w) \in E $ for any vertex $w$ other than $u$ and $v$, then we can also exchange $\kappa_u \leftrightarrow \kappa_v$ and $\theta_u \leftrightarrow \theta_v$ to preserve the likelihood.

The automorphisms cannot easily be avoided in the model and while the restriction on the highest-degree vertices helps with sampling, it does not guarantee an optimal compatibility between the embeddings.
For instance, fixing $\theta_u=0$ implies no uncertainty on $\theta_u$ and restricting some $v$ to $\theta_v\in [0,\pi]$ can affect $|V|-2$ vertices instead of only $\theta_v$.
To have the most coherent embedding sample, we opt for a post-processing alignment where we minimize the sum of squared angular separations to some arbitrary reference embedding over the automorphisms found using Nauty~\cite{mckay_2014_PracticalGraphIsomorphism}, the two possible reflections (reflection or no reflection) and the rotations $[0, 2\pi)$.

In what follows, we will compare the embeddings obtained using this Bayesian framework with the ones obtained using Mercator~\cite{garcia-perez_2019_MercatorUncoveringFaithful}---a coordinate ascent algorithm that maximizes the Eq.~\eqref{eq:likelihood}.
When embedding synthetic graphs generated with the $\mathbb{S}^1$ model, we will also use the original input parameters $(\theta, \kappa, \beta)$ as the ``ground truth'' embedding.

\subsection{Sampling}
\label{sec:sampling}

With the model in place, we can now generate embeddings by calculating expectations over the posterior density of Eq.~\eqref{eq:bayesian_model}.
For example, the expected position of a vertex can be calculated as
\begin{equation}
    \label{eq:example_integral}
    \mathbb{E}[\Theta_u | \mathcal{G}=G] = \int \theta_u\, p(\theta, \kappa,\beta|G)\, \dd \theta\, \dd \kappa\, \dd \beta,
\end{equation}
where $\Theta_u$ is the random variable of the angular position of vertex~$u$, and similar integrals can be defined for all quantities of interests, like confidence interval for the positions.
However, these integrals have, as far as we can tell, no closed-form solution, and we thus turn to sampling approximations computed using Markov Chain Monte Carlo (MCMC) algorithms.
These algorithms tend to be slow (see Supplementary Note~1 for a discussion), but we show in the Multimodality section below that there is evidence for their necessity to adequately tackle this problem.

Unfortunately, we have found that standard MCMC algorithms are not well-adapted to the $\mathbb{S}^1$ model.
Indeed, for most graphs $G$, the posterior distribution has a complicated geometry defined by several steep barriers that separate the posterior distribution in a multitude of small regions.
This is the case even for a small graph of $30$ vertices, as Fig.~\ref{fig:vertex_loglikelihood} shows: Changing the embedding coordinate of a single vertex reveals multiple local maxima of the likelihood (and thus posterior density).
The model predicts that two vertices with nearby embedding coordinates should be connected with high probability.
In the extreme case where two disconnected vertices have the same angular coordinate, the likelihood approaches $0$, causing the sharp dips shown in the Figure.
Furthermore, the gradients are undefined due to their dependence on the distance function in Eq.~\eqref{eq:edgeprob}.
Since the posterior geometry is filled with these barriers for sparse and large graphs (the number of modes increases with the number of disconnected pairs of vertices), hyperbolic embedding can be challenging for standard MCMC algorithms.

In what follows, we first shows that both a simple random walk algorithm and a sophisticated gradient-based algorithm cannot adequately sample the posterior of a small graph as a result.
Then, we demonstrate that our proposed algorithm, BIGUE, can avoid this problem by supplementing cluster-based transformations to a random walk.

\begin{figure}
    \centering
    \includegraphics{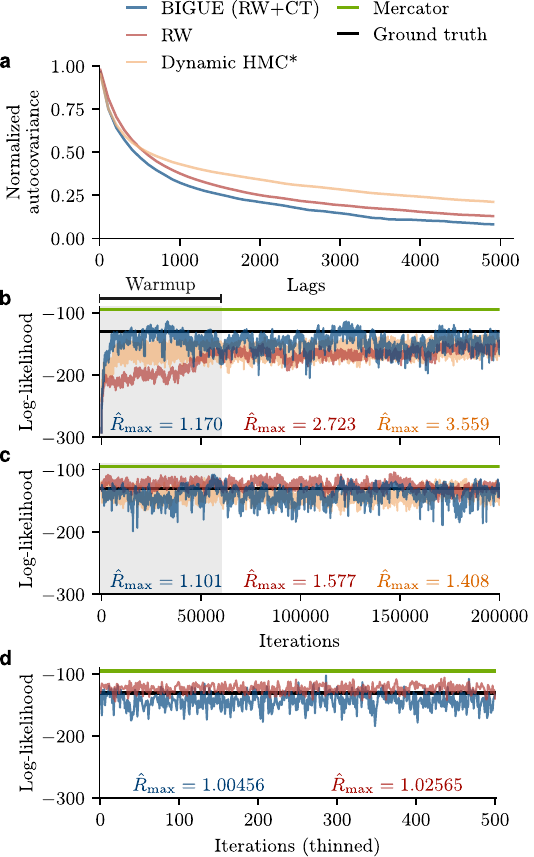}
    \caption{
        Statistics of Markov chain Monte Carlo algorithms when embedding the synthetic graph of Fig.~\ref{fig:probabilistic_embedding}.
        Results for the random walk algorithm (RW) and for our method (BIGUE) that uses both cluster transformations and a random walk (RW+CT) are displayed in red and blue, respectively.
        The orange curves show the results for dynamic Hamiltonian Monte Carlo (HMC) for the differentiable $\mathbb{S}^1$ model described in Supplementary Note~3.
        The green line displays the maximum log-likelihood obtained from Mercator after $10$ runs, each with $10$ refinements~\cite{garcia-perez_2019_MercatorUncoveringFaithful} and the black line is the ground truth embedding's log-likelihood.
        \textbf{(a)} Normalized autocovariance averaged over all parameters and chains at different lags.
        \textbf{(b-c)} Traceplot of log-likelihood of a simulated Markov chain initialized without (panel b) and with (panel c) access to ground truth information.
        When the ground truth is unknown, we initialize $\kappa_u=\deg(u)$ and draw the other parameters from their priors.
        \textbf{(d)} Traceplot of log-likelihood of a Markov chain after thinning the chain shown in panel b.
        The blue line of this panel is the sample shown in Fig.~\ref{fig:probabilistic_embedding}.
        In each case, we compute $\hat{R}_\text{max}$, the maximum potential scale reduction factor for all parameters after the iterations removed from the warm-up (grey part of traceplots)---values close to 1 are desirable.
        Four independent chains were simulated to compute $\hat{R}$ (shown on each panel with the color corresponding to the sampling algorithm) and the autocovariance, but only one is displayed (representative of the four).
    }
    \label{fig:alg_comparison_30v}
\end{figure}

\subsubsection{Baseline algorithms}

We first tested a naive random walk MCMC (RW) algorithm (details in Supplementary Methods~1) on a small synthetic graph with known ground truth embedding coordinates (shown in Fig.~\ref{fig:probabilistic_embedding}).
We quantified our results using the normalized autocovariance, the effective number of samples $S_{\mathrm{eff}}$, and $\hat{R}$---a measure of mixing (see Supplementary Methods~2 for details).
Furthermore, we tested convergence by initializing the algorithm (i) at the ground truth and (ii) with a simple initialization strategy that can be applied to observational data.
Figure~\ref{fig:alg_comparison_30v} illustrates that the RW has a high normalized autocovariance and it very slowly reaches the typical set---the part of the distribution that holds a significant probability mass and where good algorithms should naturally spend most of their time ~\cite{betancourt_2018_ConceptualIntroductionHamiltonian}.

For our second test, we wrote the model in Stan, a probabilistic programming language that implements a dynamic Hamiltonian Monte Carlo (HMC) algorithm~\cite{carpenter_2017_StanProbabilisticProgramming} for sampling from arbitrary Bayesian models.
Dynamic HMC runs slowly because each sampling iteration involves integrating a large system of differential equations.
Nonetheless, the promise of high-quality samples makes this option worth exploring.

Unfortunately, the barriers shown in Fig~\ref{fig:vertex_loglikelihood} are problematic for a purely gradient-based method such as HMC, and Stan indeed reports gradient divergences, which signal incorrect sampling.
This issue can be partially mitigated by using an approximate likelihood (described in Supplementary Note~3) that smoothens these barriers at the cost of a slightly distorted posterior.
However, even with this modification, the algorithm is unable to explore the sampling space sufficiently fast (we reach the maximum tree depth easily~\cite{carpenter_2017_StanProbabilisticProgramming}).

Stan's default parameters can be modified to allow for better exploration (increased tree depth), but even then, a critical problem remains and seems insurmountable: after a large number of iterations, the sampled distribution differs depending on the initialization, a signal that the mixing is poor; see Fig.~\ref{fig:alg_comparison_30v}.
Initializations at the ground truth yield samples of higher quality than the RW, but random initializations do not converge to the typical set even after $200,000$ iterations.
This mixing issue is likely caused by the complicated and non-convex geometry of the posterior distribution, which motivates the modifications we propose next.

\begin{figure}
    \centering
    \includegraphics[width=\columnwidth]{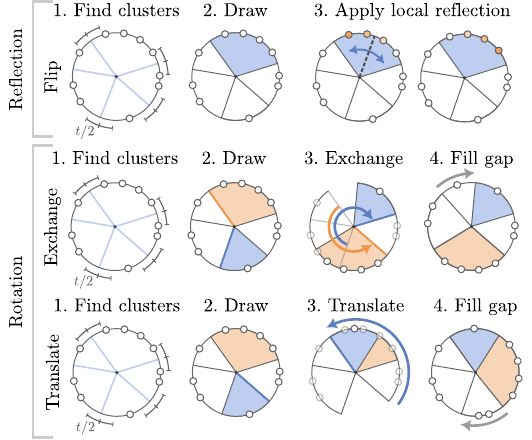}
    \caption{Summary of the cluster transformations used in BIGUE. The flip transformation targets the reflection symmetries, and the exchange and translate transformations target the rotation symmetries.}
    \label{fig:cluster_moves}
\end{figure}

\subsubsection{Cluster transformation MCMC}

We can improve upon dynamic HMC and RW by noticing that the connection probability is a decreasing function of the angular separation, such that groups of vertices at a large enough distance from one another are essentially independent.
As a result, each region is akin to a smaller and simpler embedding problem, which is the basis for the approach of Wang et al.~\cite{wang_2019_FastHyperbolicMapping, wang_2016_HyperbolicMappingComplex}.
(This is also the basis for the idea of renormalization in the $\mathbb{H}^2$ hyperbolic model in which groups of vertices are coarse-grained while partially preserving the structural integrity of the graph~\cite{garcia-perez_2018_MultiscaleUnfoldingReal, boguna_2021_NetworkGeometry}.)
Recalling that the likelihood is invariant to reflections and rotations, we propose to apply these transformations at the level of groups or clusters of vertices; c.f. Fig.~\ref{fig:cluster_moves}.

The first step for these transformations is to partition vertices into roughly independent clusters given the angular positions.
This is done by grouping vertices with an angular separation below some randomly sampled threshold in clusters.
We use this naive partitioning scheme instead of an optimization procedure (e.g. modularity maximization~\cite{patania_2023_ExactRapidLinear}) for technical reasons discussed in Supplementary Methods~1.

Once we have selected clusters, we apply one of the following transformations.
The first transformation, named flip, applies a reflection on the vertices of the selected cluster.
The second transformation, exchange, swaps the relative positions of two selected clusters.
The third transformation, translate, moves the first selected cluster to the relative position of the second cluster.
The first transformation explores local reflection symmetries, while the two others explore rotation symmetries.
In each case we select the involved cluster(s) uniformly at random from all clusters.

Our proposed algorithm, BIGUE, combines these random cluster transformations (CT) with the random walk (RW) baseline.
Cluster transformations explore the mesoscale structure of the embedding, while the random walk fine-tunes the embedding coordinates, including the parameters $\kappa$.
Each sampling iteration is either a randomly selected CT or a RW, and we calibrate the transition probabilities to the posterior distribution using the Metropolis-Hastings algorithm; see Supplementary Methods~1 for the technical details.
(One could also combine random cluster transformations with dynamic HMC, but the high computational cost of dynamic HMC turns out to outweigh its benefits. A naive random walk provides cheaper and sufficient fine-tuning in our experiments.)

Figure~\ref{fig:alg_comparison_30v} shows that BIGUE has the lowest autocovariance and reaches the typical set faster the other algorithms.
The fluctuations at equilibrium are also larger, which suggests that it explores the posterior more efficiently.
Furthermore, BIGUE's maximum potential reduction factor $\hat{R}_\text{max}$ is the lowest, suggesting that it has better mixing.
However, the three algorithms yield $\hat{R}$ values above the recommended $1.01$ threshold~\cite{vehtari_2021_RankNormalizationFoldingLocalization}.

A direct fix to high normalized autocovariance and potential reduction factor is to thin the chains, that is, only use every $k$-th sample point of the chain.
Guided by Fig.~\ref{fig:alg_comparison_30v}a, we set $k=10,000$ for BIGUE and the RW and obtain the chains of panel d.
We now see that BIGUE's chains have $\hat{R}_\text{max}<1.01$ while those of the RW do not.
Moreover, the fluctuations of the RW are still smaller, indicating, again, that the sampling space is better explored by BIGUE.

From the samples of Fig.~\ref{fig:alg_comparison_30v}, we also compute the effective sample size of the chains, which is related to the estimation error in the Markov Chain central limit theorem.
The median effective sample sizes $S_\text{eff}$ across all parameters were, for BIGUE: $486$, $462$, and  $1814$ (in panels b,c, and d respectively).
For the random walk MCMC, we obtained medians of $S_\text{eff}=248$, $384$, and $1390$, and for dynamic HMC, they were equal to $593$ and~$347$.
This provides further evidence that cluster transformation benefit the random walk algorithm.
Autocovariance, $\hat{R}$ and $S_\text{eff}$ and their adaptation to periodic random variables are explained in Supplementary Methods~2.

\begin{figure}
    \centering
    \includegraphics{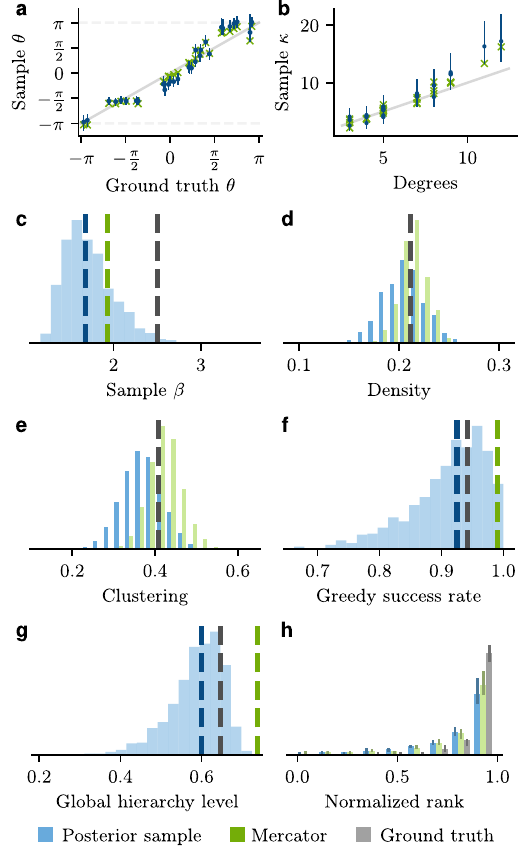}
    \caption{
        Posterior estimates and posterior predictive distribution for the synthetic graph used in Fig.~\ref{fig:alg_comparison_30v}.
        \textbf{(a)} Angular coordinates $\theta$.
        \textbf{(b)} Parameters $\kappa$.
        \textbf{(c)} Inverse temperature $\beta$.
        \textbf{(d)} Density.
        \textbf{(e)} Clustering (the transitivity).
        \textbf{(f)} Greedy routing success rate.
        \textbf{(g)} Global hierarchy level~\cite{garcia-perez_2016_HiddenHyperbolicGeometry}.
        \textbf{(h)} Normalized rank of the removed edges' existence probability among all the unconnected pairs (see below).
        In each case, $2000$ total embeddings were sampled from four independent chains.
        All shades of blue, green and gray display the values for the posterior sample, Mercator and the ground truth, respectively.
        In panels (a,b), points denote the median, and the error bars cover the interquartile range.
        In panels (c-g), the vertical dotted lines show point estimates computed with the Mercator and ground truth embeddings.
        In panels (d,e), we generated graph samples for Mercator and the ground truth by conditioning the model's likelihood on point estimates of the embedding.
        In panels (c, f,g), the blue dashed line is the median of the posterior sample.
        In panel (h), we test the algorithm on a link prediction task~\cite{kumar_2020_LinkPredictionTechniques}, and report the rank of removed edges as predicted by the likelihood. Pairs of embedding and graphs of 30 vertices were sampled from the prior. For each graph, $5\%$ of the edges were then randomly removed ($4$ to $5$ edges), and the normalized ranks of removed edges were calculated and binned.  (Disconnected graphs were rejected for compatibility with Mercator).
        The figure reports the median and interquartile range of the normalized rank frequencies calculated across graphs, removals, and samples.
        The AUC values are shown in Supplementary Note~5.
    }
    \label{fig:error_bars}
\end{figure}

In light of these simulations, we conclude that BIGUE outperforms both dynamic HMC and the RW in all metrics, and that in general, we expect these algorithms to perform equivalently to BIGUE at best.
For this reason, the remainder of the paper uses BIGUE to sample the posterior distribution.

It may seem surprising that BIGUE relies on clusters in the embedding while the vertices are uniformly distributed according to prior density of the generative model.
However, the clusters found need not be well separated to be useful in sampling.
This is due to the locality of connections: If some pairs of vertices are sufficiently distant, then we can reasonably attempt to separate these vertices as they might not be connected anyway.
Conversely, vertices that are closeby in the best embedding will tend to be connected and can thus be moved as a unit instead of one at a time.

We note that Fig.~\ref{fig:alg_comparison_30v} highlights one counterintuitive aspect of continuous random variables~\cite{betancourt_2018_ConceptualIntroductionHamiltonian}: The neighborhood of the likelihood maximum accounts for a small fraction of the probability mass because probabilities are obtained through integration with respect to the Lebesgue measure.
While the log-likelihood maximum (found by Mercator) has a high probability density, its surroundings cover a larger volume (the typical set) and hold a much larger probability mass.
Hence, our MCMC sample concentrates away from the maximum likelihood, as expected.
(That said, we note that cluster transformations could easily be added to maximization routines if a point estimate was the goal, something we leave for future work.)

\section{Results}
\label{sec:uncertain_embeddings}

Having a reliable algorithm to sample embeddings from the posterior distribution, we can now estimate embeddings and integrals like Eq.~\eqref{eq:example_integral}.
This section illustrates how the resulting Bayesian approach compares to maximum likelihood estimation both on a synthetic graph and various empirical networks.

\subsection{Embedding and network properties}

\begin{figure}
    \centering
    \includegraphics{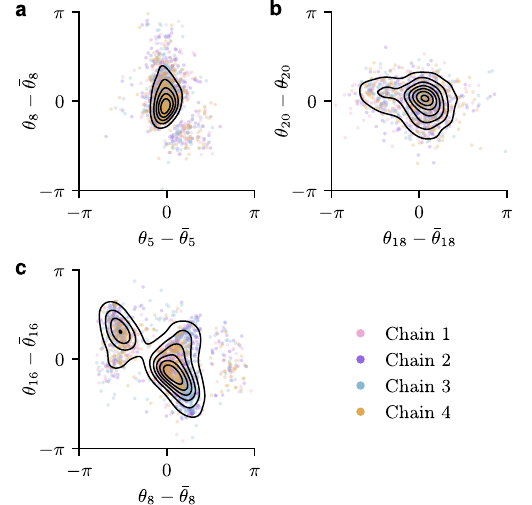}
    \caption{Multimodality and non-normality of the posterior distribution, shown with the marginal posterior distributions of pairs of angular coordinates in the embedding of \textbf{(a)} the synthetic graph in Fig.~\ref{fig:alg_comparison_30v} \textbf{(b)} Zachary's karate club~\cite{zachary_1977_InformationFlowModel} \textbf{(c)} a synthetic graph of $30$ vertices with two conflicting embeddings, as described in section~\ref{ssec:multimodality} ($\theta_0$ has two different ground truth positions).
    Samples are obtained with the thinned BIGUE algorithm, and chains comprise $500$ random embeddings.
    The color of each circle indicates from which chain it was sampled.
    The black lines delineate the density of a normal kernel density estimation of the marginal.
    The sample averages $\bar{\theta}_u$ are circular averages as described in Supplementary Methods~2.
    }
    \label{fig:marginal_pair_theta}
\end{figure}

\begin{figure*}
    \centering
    \includegraphics{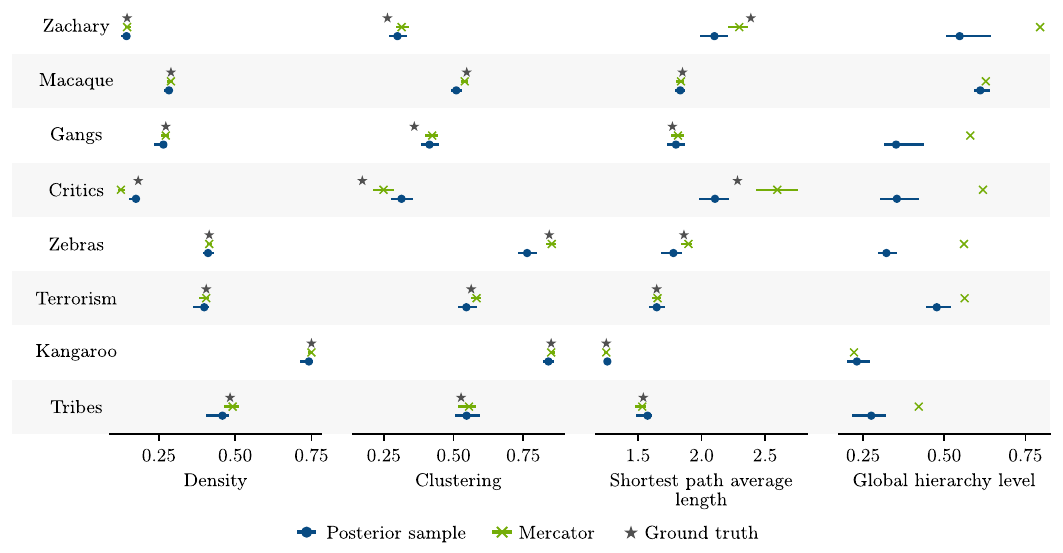}
    \caption{Comparison of network embeddings obtained from a maximum likelihood estimator (Mercator, green cross) and the posterior of a Bayesian model (BIGUE, blue circle) for various empirical networks. These networks are Zachary's karate club~\cite{zachary_1977_InformationFlowModel}, cortical connectivity of the macaque~\cite{young_1997_OrganizationNeuralSystems}, street gangs in Montreal~\cite{descormiers_2011_AlliancesConflictsContradictions}, dutch literary critics~\cite{denooy_1999_LiteraryPlaygroundLiterary}, zebra social interactions~\cite{sundaresan_2007_NetworkMetricsReveal}, connections in greek terrorist group~\cite{rhodes_2009_InferringMissingLinks}, kangaroo dominance relationships~\cite{grant_1973_DominanceAssociationMembers} and New Guinea tribes friendships~\cite{read_1954_CulturesCentralHighlands}.
    The value for the original network are displayed in gray stars when available.
    Vertices outside the largest connected component were removed in order to run Mercator.
    The horizontal bars show the 50\% highest density intervals, and points within are the median.
    Supplementary Table~1 reports the values for this figure.
    }
    \label{fig:empicial_networks}
\end{figure*}

We first compare our Bayesian model to Mercator on the synthetic graph used in Fig.~\ref{fig:alg_comparison_30v}.
Figure~\ref{fig:error_bars} shows what such an analysis might look like, using credible intervals for the embedding coordinates and their transformations, computed with samples drawn from the chains depicted in Fig.~\ref{fig:alg_comparison_30v}d.

We see in Fig.~\ref{fig:error_bars}a that Mercator's estimation generally lies within the estimated credible intervals, revealing some leeway in the coordinates compatible with the data $G$.
The ground truth value is sometimes not within the interval because the graph is small (see Supplementary Figure~1, where the graph contains $100$ vertices, and the coordinates are recovered almost perfectly).
As expected, for most vertices $u$, the $\kappa_u$ parameters concentrate on the vertices' original degrees (Fig.~\ref{fig:error_bars}b), but there are discrepancies.
Note that the relationship $\mathbb{E}[K | \kappa_u] =\deg(u)$, mentioned in the Model definition section, is valid assuming an infinite number of vertices.
This is why, for a high-degree vertex $u$ in a small graph, $\kappa_u$ is larger to compensate for the fact that there is a finite number of vertices.
Figure~\ref{fig:error_bars}c also shows that the posterior covers the original $\beta$ value, although it is in a somewhat low probability region.
The posterior distribution agrees with Mercator and places a higher likelihood than the ground truth on $\beta$ being a bit under $2$---this is due to random fluctuations in the instantiation of $G$ conditioned on the ground truth coordinates.

Next, we turn to the properties of the graph.
The edge probabilities of good embeddings should yield graphs with properties similar to those of the original graph.
As a result, if the embedding is accurate, we expect that the graphs generated with estimated coordinates will be similar to the original graph.
We first verify this with the maximum likelihood found with Mercator; the resulting distribution of graph properties is centered around the observed graph in Fig.~\ref{fig:error_bars}d,e.
We note that although Mercator yields a point-wise estimation, different graphs can be sampled from the same embedding, which explains why there are many density and clustering values for Mercator.
With BIGUE, each sample point corresponds to a random graph using the likelihood, and consequently, we can marginalize the graph properties over a much larger parameter space.
(Formally, we draw from the posterior predictive distribution of the model.)
The resulting distributions are wider and now capture parameter uncertainty; see Fig.~\ref{fig:error_bars}d,e.

The embeddings also capture less obvious properties.
For example, hyperbolic embeddings can play a role in finding effective and efficient greedy routing on the network.
In the greedy routing algorithm~\cite{krioukov_2010_HyperbolicGeometryComplex}, hyperbolic coordinates act as addresses, and one attempts to reach the target $v$ from a source $u$ by repeatedly following the edge that leads to the neighbor closest to the destination $u$ (in hyperbolic space).
The main issue with this algorithm is that paths can sometimes devolve into ``greedy loops'' that lead nowhere.
The greedy success rate evaluates the extent to which this is an issue as the proportion of greedy routes that successfully reach their destination.
The posterior samples allows us to compute error bars on the greedy success rate, but also to comment the navigability of the ensemble of plausible embeddings.
For instance, Fig.~\ref{fig:error_bars}f shows that, while many embeddings are as navigable as the embedding inferred by Mercator, the majority of plausible embeddings are less reliable.

The hierarchical organization of graphs is another example of complex properties captured by hyperbolic embeddings.
The global hierarchy level was introduced to quantify this, using a function of the angular distance between neighboring vertices with coordinates in the outer and inner parts of the embedding~\cite{garcia-perez_2016_HiddenHyperbolicGeometry}.
Figure~\ref{fig:error_bars}g illustrates how, for the synthetic graph analyzed, most plausible embeddings have smaller global hierarchy levels than both the ground truth and the Mercator embeddings.

Finally, embeddings for the $\mathbb{S}^1$ model can be viewed as classifiers for edges since the likelihood maps potential edges (binary classes) to edge probabilities.
If the embedding is a good representation of the graph, edges should have a high probability, while non-edges should have a low probability.
We test the performance in the edge prediction task by generating many embeddings and synthetic graphs of 30 vertices and removing 5\% of its edges.
We then compute the rank of the removed edges among all the unconnected pairs, using their inferred connection probability, Eq.~\eqref{eq:edgeprob}.
If the removed edges still have a relative high existence probability, the embedding is a good predictor and the ranks should approach $1$.
Figure~\ref{fig:error_bars}h shows that Mercator and BIGUE perform almost identically and, unsurprisingly, the original embedding performs better.
This suggests that both embeddings are robust to small perturbations.

\subsection{Multimodality}
\label{ssec:multimodality}

Next, we inspect the properties of the posterior distribution of coordinates in more detail.
MCMC algorithms are computationally expensive, and one might wonder if we could do away with sampling altogether---be it HMC, BIGUE, or another algorithm---by quantifying uncertainty with Laplace's approximation or a variational approach~\cite{blei2017variational} instead.
These approximations are attractive because they yield error bars for all quantities nearly for free, and they have strong theoretical backing, e.g., the Bernstein–von Mises theorem that guarantees convergence to the approximated distribution in the big data regime under certain conditions.
Unfortunately, for small graphs and $\mathbb{S}^1$, at the very least, we find that the posterior distributions are not multivariate normal distributions and that MCMC is truly necessary.

Treating our sample as truly representative of the posterior distribution, we computed the p-values of the Henze-Zirkler normality test~\cite{henze_1990_ClassInvariantConsistent} for all pairs of parameters.
This test compares the empirical characteristic function to its pointwise limit under the null hypothesis that the data follows a multivariate normal distribution.
We found the largest p-value for this test to be $1.11 \times 10^{-25}$ for the example synthetic graph.
This is not surprising when looking at the marginal distribution of the pair of coordinates shown in Fig.~\ref{fig:marginal_pair_theta}a: The bulk of the distribution is skewed, and there is a region of very low probability mass next to the bulk.
All marginalizations of a multivariate normal yield another multivariate normal; thus, finding one non-normal marginal is enough to reject it as a model for the whole.

It could be that only synthetic graphs show non-normal posterior distributions, but it turns out that the embedding of well-known graphs not explicitly generated by the model behaves the same way.
For example, the posterior of Zachary's Karate Club is also not normal: one pair of parameters has a p-value of $0.092$ for the Henze-Zirkler normality test, but the complete posterior has a p-value of $0$ numerically; Fig.~\ref{fig:marginal_pair_theta}b shows that joint marginals for this graph are not ellipsoids.

One of the most compelling pieces of evidence for the non-normality of hyperbolic embeddings, is the ease with which it can be induced using a mixture for the embedding coordinates.
A simple procedure that achieves multimodality is as follows.
First, we give random coordinates to the vertices as usual.
But when we generate the edges of $G$, we use either the original or an updated angular coordinate for a single vertex $u$ (e.g., $\theta_u\mapsto\theta_u + 2 \mod 2\pi$), with probability $0.5$, thereby connecting $u$ to two incompatible sets of vertices. Figure~\ref{fig:marginal_pair_theta}c shows the marginal distribution of two angular positions for the resulting graph.
Neither of these two chosen vertices is the vertex with a superposition of angular coordinates, yet the marginal is clearly multimodal.
In our experiments, we also found that almost all joint marginals that include at least one angular coordinate are bimodal; a single vertex with an ambiguous position is sufficient to induce the behavior.

\subsection{Observational study}

To conclude our analysis, we embed a collection of empirical networks with BIGUE and contrast our results with Mercator's; the experiments are summarized in Fig.~\ref{fig:empicial_networks} (see Appendix for a table giving numerical values).
Both algorithms reproduce the observed density and clustering (transitivity) of the graph.
The shortest paths average length is also reproduced by both algorithms except for the Zachary's karate club and the Dutch literary critics networks, where both algorithms fail.
In these cases, it is likely that the $\mathbb{S}^1$ model cannot offer a good representation altogether.
And, as we have argued above, Mercator yields a single embedding and thus ignores parameter uncertainty, leading to tighter estimated intervals.

Properties that only depend on coordinates, like the global hierarchy level, have no associated ground truth in observational data and can only be compared across different embedding algorithms.
Except for the macaque neural network, we find that Mercator systematically yields higher values of the hierarchy than BIGUE's, usually by a large margin.
This suggests that graphs are less hierarchical than they might seem if we only had considered the Mercator embedding.

\section{Conclusions}

Hyperbolic random graphs are unquestionably useful and can explain many properties observed in empirical networks, but existing estimation algorithms neglect the significant uncertainty and multimodality that can be present in these models, while off-the-shelf uncertainty quantification methods fall short.
We provided evidence that the dynamic HMC is unable to sample the posterior of a Bayesian hyperbolic embedding model, and that supplementing random cluster transformations to a generic random walk is sufficient to sample embeddings of small graphs.
While maximum likelihood estimators such as Mercator overfit the coordinates and do not provide error bars on embedding, our algorithm BIGUE provides an accurate posterior sample for the error bars on the graph and embedding properties.

As future work, we believe that generalizing cluster transformation on the $D+1$-sphere could be a promising avenue~\cite{jankowski_2023_DMercatorMethodMultidimensional}.
While each additional dimension would increase the computational cost by adding another parameter per vertex, this tradeoff may be worthwhile.
The higher-dimensional space would allow vertices to navigate around each other more freely, potentially avoiding the barriers illustrated in Fig.~\ref{fig:vertex_loglikelihood}.

Another challenge we leave for future work is to improve the algorithm's overall computational efficiency.
Currently, BIGUE's main limitation is its poor scaling with vertex count---it struggles to properly sample graphs with more than 100 vertices in a reasonable timeframe.
While basic optimizations like GPU-accelerated likelihood calculations and a more efficient programming language would help, the most promising speedups may lie in relying on HMC for the adjustment of positions.
Naively using external implementations turned out to be too costly, but an integrated solution may work well.
Other potential speed improvements include using negative sampling to approximate the likelihood~\cite{liu_2024_RevisitingSkipGramNegative} instead of computing the likelihood in $O(|V|^2)$ operations; sampling vertices individually; incorporating graph symmetries during sampling---by swapping affected vertex positions---or even approximate symmetries~\cite{pidnebesna_2025_ComputingApproximateGlobal}.

Beyond sampling applications, the success of cluster transformations suggests they could enhance existing likelihood maximization algorithms like Mercator.

\section*{Data availability}
The data were obtained from an online repository~\cite{peixoto_2020_netzschleuder} or can be fetched from the original papers~\cite{zachary_1977_InformationFlowModel, young_1997_OrganizationNeuralSystems, descormiers_2011_AlliancesConflictsContradictions,denooy_1999_LiteraryPlaygroundLiterary, sundaresan_2007_NetworkMetricsReveal, rhodes_2009_InferringMissingLinks, grant_1973_DominanceAssociationMembers, read_1954_CulturesCentralHighlands}.
All the data are also available with the code at \url{https://doi.org/10.5281/zenodo.15272626}.

\section*{Code availability}

A Python implementation of BIGUE is available at \url{https://doi.org/10.5281/zenodo.15189658}.
The Python code for the numerical analysis of this paper is available at \url{https://doi.org/10.5281/zenodo.15272626}.

\section*{Acknowledgments}

This work was supported by Conseil de recherches en sciences naturelles et en génie du Canada (AA), Sentinelle Nord (SL, AA) and the Fonds de recherche du Québec (SL), and the Vermont Complex Systems Institute (JGY).
We acknowledge Calcul Qu\'ebec and Alliance de recherche numérique du Canada for their technical support and computing infrastructures.

\section*{Author contributions}

S.L. designed the algorithm, implemented the method and wrote the paper.
A.A and J.G.Y supervised the work, and contributed to the research design, result analysis, and paper writing.

\section*{Competing interests}
The authors declare no competing interests.

\medskip

\end{document}